\documentclass[traditabstract]{aa}
\usepackage{graphicx}
\usepackage{txfonts}
\usepackage[authoryear]{natbib}
\begin{document}

\title{Non Thermal Support for the Outer Intracluster Medium}

\author{A. Cavaliere\inst{1}, A. Lapi\inst{1,2}, R. Fusco-Femiano\inst{3}}
\institute{$^1$ Dip. Fisica, Univ. `Tor Vergata', Via Ricerca Scientifica 1, 00133
Roma, Italy.\\
$^2$ SISSA, Via Bonomea 265, 34136 Trieste, Italy.\\
$^3$ INAF-IASF, Via Fosso del Cavaliere, 00133 Roma, Italy.}

\date{\today}

\abstract{We submit that non thermalized support for the outer intracluster 
medium in relaxed galaxy clusters is provided by turbulence, driven by 
inflows of intergalactic gas across the virial accretion shocks. We expect 
this component to increase briskly during the cluster development for 
$z\lesssim 1/2$, due to three factors. First, the accretion rates of gas and 
dark matter subside, when they feed on the outer wings of the initial 
perturbations in the accelerating Universe. Second, the infall speeds 
decrease across the progressively shallower gravitational potential at the 
shock position. Third, the shocks eventually weaken, and leave less thermal 
energy to feed the intracluster entropy, but relatively more bulk energy to 
drive turbulence into the outskirts. The overall outcome from these factors 
is physically modeled and analytically computed; thus we ascertain how these 
concur in setting the equilibrium of the outer intracluster medium, and  
predict how the observables in X rays and $\mu$waves are affected, so as to 
probe the development of outer turbulence over wide cluster samples. By the 
same token, we quantify the resulting negative bias to be expected in the 
total mass evaluated from X-ray measurements.} 

\keywords{galaxies: clusters: intracluster medium --- turbulence --- X-rays:
galaxies: clusters --- methods: analytical.}

\authorrunning{A. Cavaliere et al.}
\titlerunning{Non Thermal Support for the ICM}

\maketitle

\section{Introduction}

In galaxy clusters the gravitational potential wells set by Dark Matter (DM)
masses $M\sim 10^{15}\,M_{\odot}$ are filled out to the virial radius $R\sim$
Mpc by a hot thin medium at temperatures $k_BT\sim $ several keVs, with
central particle densities $n\sim 10^{-3}$ cm$^{-3}$.

Such a medium constitutes an optimal electron-proton plasma (that we 
appropriately name IntraCluster Plasma, ICP), with its huge ratio of the 
thermal to the mean electrostatic energy $k_B T/e^2\, n^{1/3}\sim 10^{12}$, 
and the relatedly large number $n \, \lambda^3_D = (k_B T/4 \pi \, e^2\, 
n^{1/3})^{3/2} \, \sim 10^{16}$ of particles in the Debye cube. It emits 
copious X-ray powers $L_X\propto n^2\, T^{1/2}\, R^3\sim 10^{45}$ erg 
s$^{-1}$ via thermal bremsstrahlung, but with long radiative cooling times 
over most of the cluster volume. 

Thus on scales longer than the electro-proton mean free path the ICP 
constitutes a quasi-neutral, simple fluid with $3$ degrees of freedom in
thermal equilibrium, and with effective particle mass $\mu m_p\approx 
0.6\,m_p$ in terms of the proton's $m_p$. This affords precision modeling on 
the radial scale $r$ $-$ to begin with $-$ for the distributions of density 
$n(r)$ and temperature $T(r)$, as to match the wealth of current and upcoming 
data concerning the emissions in X rays \citep[e.g., reviews 
by][]{Snowden2008, Giacconi2009}, and the strengths $y \propto n\, T_e $ of 
the Sunyaev-Zel'dovich (1972) scattering in $\mu$waves 
\citep[e.g.,][]{Birkinshaw2007, Schafer2007}. 

In fact, simple yet precise modeling is provided by the Supermodel 
\citep[SM;][]{Cavaliere2009}. This is based on the run of the ICP specific
`entropy' (adiabat) $k\equiv k_B T/n^{2/3}$ set by the processes for its 
production. 

The entropy is raised at the cluster centers due to the energy discharged by 
AGN outbursts \citep[see][]{Valageas1999, Wu2000, Cavaliere2002, 
McNamara2007}, and by deep mergers \citep[see][]{McCarthy2007, 
Markevitch2007} often followed by inner sloshing 
\citep[see][]{Zuhone2010}. At the other end, much entropy is continuously 
produced at the virial boundary; there the ICP is shocked by the supersonic 
gravitational inflow of gas accreted from the environment along with the DM 
\citep[see][]{Tozzi2001, Voit2005, Lapi2005}, and is adiabatically stratified 
into the DM potential well.

These physical processes concur to originate ICP entropy distributions with 
spherically averaged profiles $k(r)=k_c+(k_R-k_c)\,(r/R)^a$; these comprise a 
central \textit{floor} $k_c\approx 10-100$ keV cm$^2$, and an outer ramp with 
\textit{slope} around $a\approx 1$, rising to adjoin the \textit{boundary} 
values $k_R \sim$ some $10^3$ keV cm$^2$. Such values and shapes are 
consistent with recent observational analyses by \citet{Cavagnolo2009} and 
\citet{Pratt2010} out to $r \approx R/2$\footnote{Note that $R/2\approx 
R_{500}\approx 2\, R_{200}/3$ holds in terms of the radii inside which the 
average DM overdensity relative to the critical universe amounts to $500$ and 
$200$, respectively.}. 

The ensuing gradient of the thermal pressure $p(r)\propto k(r)\,n^{5/3}(r)$ 
is used in the SM to balance the DM gravitational pull $-G\,M(<r)/r^2$ and 
sustain hydrostatic equilibrium (HE) out to the virial boundary. The 
HE equation may be written as $\mathrm{d}\ln T/\mathrm{d}\ln r =3\,a/5-2\, 
b/5$ in terms of the entropy slope $a(r)\equiv \mathrm{d}\ln k/\mathrm{d}\ln 
r$, and of the potential to thermal energy ratio $b(r)\equiv \mu m_p\, 
v_c^2/k_B T$ with $v_c^2\equiv G\, M(<r)/r$. Whence we directly derive the 
temperature profile \citep[see][their Eq.~7]{Cavaliere2009} 
\begin{equation}
{T(r)\over T_R}=\left[{k(r)\over k_R}\right]^{3/5} \left\{1 + {2\over 5}\, 
b_R\, \int^R_r{\mathrm{d} x\over x}\,{v_c^2(x)\over v_R^2}\, \left[{k_R\over 
k(x)}\right]^{3/5}\right\}~ 
\end{equation}
in terms of the entropy run $k(r)$ and the boundary values at $r = R$. The 
density follows $n(r)=[k_B T(r)/k(r)]^{3/2}$, so that $T(r)$ and $n(r)$ are 
\emph{linked}, rather than independently rendered with 
multi-parametric expressions as in other approaches. From $T(r)$ and $n(r)$ 
the X-ray and SZ observables are readily derived and compared with data. 

In preparation to developments given below, we stress that the few parameters 
specifying $k(r)$ are enough for the SM to provide remarkably good fits to 
the detailed X-ray data on surface brightness and on temperature profiles of 
many clusters \citep[see][]{Fusco2009}. These include central temperature 
profiles of both main classes identified by \citet{Molendi2001}: the 
cool-cored CCs with a central dip, and the centrally flat, non-cool-cored 
NCCs. The SM intrinsically links these morphologies to low or high 
entropy levels $k_c\sim 10^1$ or $\sim 10^2$ keV cm$^2$, respectively; these 
are conceivably imprinted by energy inputs from central AGN outbursts or from 
deep mergers (\citealp{Cavaliere2009}, \citealp{Fusco2009}).

The SM also covers diverse outer behaviors including cases where the entropy 
production decreases and its slope $a$ abates \citep[see][and data references 
therein]{Lapi2010}, to the effect of producing steep temperature profiles. 
The interested reader may try for her/himself other clusters on using the 
fast SM algorithm made available at the website 
\textsl{http://people.sissa.it/$\sim$lapi/Supermodel/.} Here we pursue 
another consequence of the diminishing entropy production, namely,  
turbulence arising in the outskirts of \emph{relaxed} CC clusters. 

A number of observations (in particular with the \textsl{Suzaku} satellite, 
see \citealp{George2009}, \citealp{Bautz2009}, \citealp{Hoshino2010}, 
\citealp{Kawaharada2010}), support the notion that HE may be contributed by 
non-thermalized, turbulent motions occurring on scales of several $10^2$ kpc 
inwards of $R$. On the other hand, several simulations resolve a 
variety of shocks in and around clusters (e.g., \citealp{Ricker2001}, 
\citealp{Markevitch2007}, \citealp{Skillman2008}, \citealp{Vazza2010}).

We focus on the accretion shocks that originate at the virial 
boundary from inflows of gas preheated at temperatures $10^6$ K by sources 
like outer stars and AGNs or by hydrodynamical processes like shocks around 
filaments. The above two pieces of information lead us to investigate 
whether the physics of the \emph{virial} accretion shocks indeed 
requires \emph{turbulence} to develop also in the outskirts of relaxed 
clusters under the smooth inflows that prevail there (\citealp{Fakhouri2010}, 
\citealp{Genel2010}, \citealp{Wang2010}). 

\section{Virial accretion shocks: entropy and turbulence}

The key physical agent is constituted by the residual bulk flows downstream 
the virial accretion shocks. The latter actually form a complex 
network \citep[`shock layer', see][]{Lapi2005} modulated to different 
strengths by the filamentary structure of their environment. Shock curvature 
leading to baroclinic instabilities and vortical flows, and/or sheared 
inflows inevitably arise as pioneered in the context of cosmic structures by 
\citet{Doroshkevich1973} and \citet{Binney1974}, and recently demonstrated by 
several hydro-simulations (see \citealp{Iapichino2008}, \citealp{Ryu2008}, 
\citealp{Lau2009}, \citealp{Paul2010}). 

\subsection{Physics of virial accretion shocks}

Boundary shocks of high average strength arise in conditions of intense 
inflows of outer gas with supersonic speed $v_1$ corresponding to Mach 
numbers $\mathcal{M}^2\equiv v_1^2/c_1^2\sim 10^2$ relative to the outer 
sound speed $c_1\equiv (5\,k_B T_1/3\,\mu m_p)^{1/2}$. These shocks 
effectively thermalize the inflows, to produce postshock temperatures close 
to the ceiling $k_B\, T_2 \sim \mu m_p\,v_1^2 /3$ that would mean full 
conversion of the infall $m_p\, v_1^2/2$ to thermalized energy $3\, k_B\, T/ 
2 \mu$ per electron-proton pair. But even strong shocks hovering at 
$R_s\approx R$ leave some residual postshock bulk flows with speed 
$v_2\approx v_1/4$, see \citet{Lapi2005}. These correspond to a kinetic 
energy ratio $v_2^2/v_1^2\approx 6.3\%$. 

Such a ratio is bound to grow, however, during the outskirts development. On 
the DM side, the latter develop inside-out by secular accretion after the 
early central collapse (see \citealp{Lapi2009}, \citealp{Wang2010}, and 
references therein). Such a trailing accretion feeds scantily on the outer, 
declining \emph{wings} of the initial DM density perturbation that develops 
into a cluster, and is further impaired by the cosmic expansion 
\emph{accelerated} by the Dark Energy \citep[cf.][]{Komatsu2010}. In these 
conditions, the inflows will peter out, shock thermalization will be reduced 
and eventually the shock themselves weakened, to leave postshock bulk 
energies enhanced well above the ratio $6.3\%$. 

To quantify the issue, we describe the perturbation shape in terms of the 
effective powerlaw $\delta M/M\propto M^{-\epsilon}$ that modulates the mass 
excess $\delta M$ accreting onto the current mass $M$; in particular, low 
values of the \emph{shape} parameter $\epsilon\lesssim 1$ apply to the 
perturbation body, but $\epsilon$ grows larger for the outskirts 
\citep[see][]{Lu2006}. A shell $\delta M$ will collapse on top of $M$ when 
$\delta M/M$ attains the critical threshold $1.69\, D^{-1}(t)$ in terms of 
the linear growth factor $D(t)$. So the parameter $\epsilon$ also modulates 
the average mass \emph{growth} reading $M(t)\propto D^{1/\epsilon}\propto 
t^{d/\epsilon}$, with the growth factor represented by the powerlaw 
$D(t)\propto t^{d}$ in terms of the exponent $d$; the latter decreases from 
$2/3$ to $1/2$ as the redshift lowers from values $z\ga 1$ to $z\lesssim 
1/2$, cf. \citet{Weinberg2008}. Thus the outskirts develop at accretion rates 
$\dot M/M\approx d/\epsilon\,t$ that \emph{lower} as $\epsilon$ takes on 
values exceeding $1$ in the perturbation wings (and formally diverging in 
voids), and as $d$ decreases to $1/2$ at late cosmic times.

\begin{figure}
\centering
\includegraphics[width=7cm]{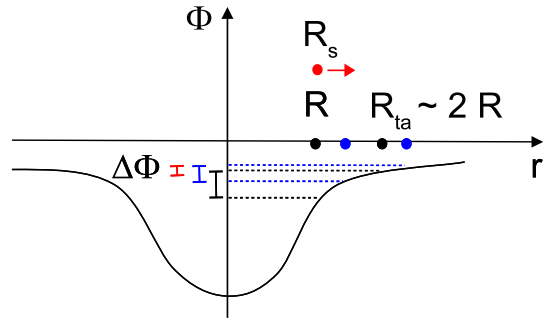}
\caption{To illustrate the gravitational potential governing gas infall. 
As the DM outskirts develop (see Sect.~2), both the virial $R$ and the 
turnaround radius $R_{\rm ta}$ shift outwards, retaining the ratio $R_{\rm 
ta}/R\approx 2$. Meanwhile, the outer potential becomes shallower and 
$\Delta\Phi$ lower from the value marked in black to that in blue. Moreover, 
the shock position $R_s$ slowly outgrows $R$, lowering yet the drop to the 
value marked in red.} 
\end{figure}

\begin{figure}
\centering
\includegraphics[width=\columnwidth]{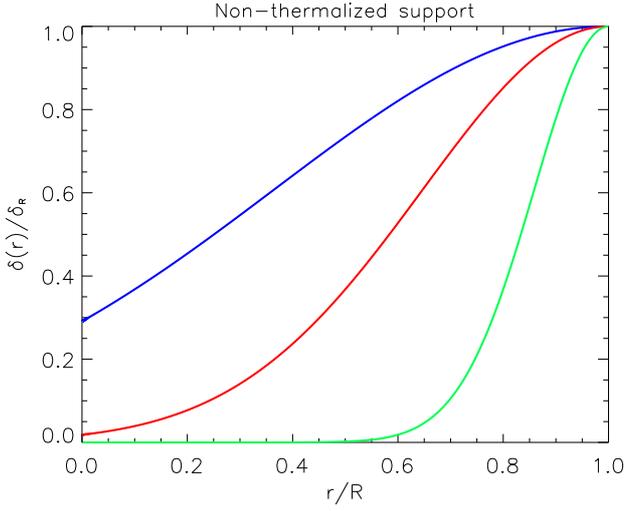}
\caption{Non-thermalized pressure support $\delta(r)\equiv p_{\rm 
nth}/p_{\rm th}$ normalized to the boundary value $\delta_R$ computed after 
Eq.~(8). The blue line refers to $\ell=0.9\, R$, red line to $\ell=0.5\, R$, 
and green line to $\ell=0.2\, R$.}
\end{figure}

On the ICP side, the outer gas will accrete with a \emph{lower} infall speed 
$v_1$. As illustrated in Fig.~1, the latter is set to $v_1^2=2\,\Delta\Phi$ 
by the outer gravitational potential drop $\Delta\Phi\equiv \int^{R_{\rm 
ta}}_{R}{\mathrm{d}r}~G\,\delta M/r^2$; this is experienced by successive 
shells of DM and gas that $-$ after an initial expansion $-$ turn around at 
the radius $R_{\rm ta}\approx 2\,R$ to begin their infall toward the shock at 
$R_s \approx R$. Thus the potential drop $\Delta\phi\equiv\Delta\Phi/v_R^2$ 
(normalized to the circular velocity $v_R^2=G\,M/R$ at $r=R$) reads 
\begin{equation}
\Delta\phi={{1-(R/R_{\rm ta})^{3\epsilon-2}}\over 3\epsilon-2}~.
\end{equation}
This is seen (cf. Fig.~1) to become \emph{shallower} during the outskirts
development as $\epsilon$ exceeds $1$; then the approximation
$\Delta\phi\approx (3\,\epsilon-2)^{-1}\approx (3\epsilon)^{-1}$ applies.

Actually, the shock position $R_s$ slowly outgrows the virial $R$ to approach
$R_{\rm ta}$, and this yields an even \emph{lower} effective potential drop
\citep[see][]{Voit2003, Lapi2005}. In fact, it can be shown that for
$\epsilon>1$ the shock position may be approximated as $R_s\approx
2\,R\,(1-4\,\epsilon^{-2})$, see \citet{Lapi2010}; so from Eq.~(2) we obtain
$\Delta\phi\approx 4\,\epsilon^{-2}$.

\begin{figure*}
\centering
\includegraphics[width=15cm]{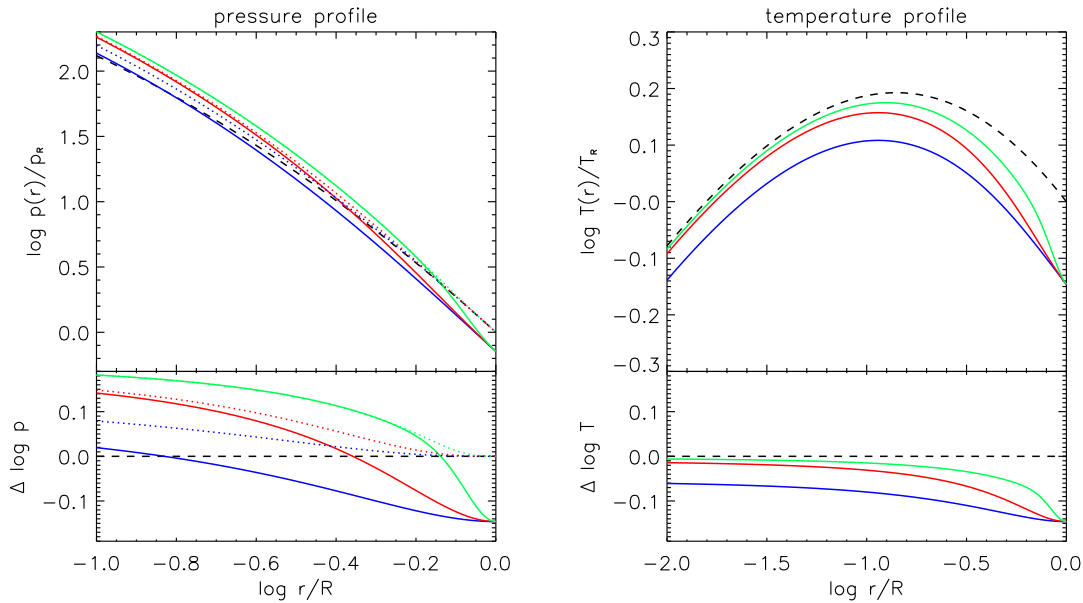}
\caption{Profiles of pressure and temperature computed with the SM. The 
dashed line illustrates the pure thermal (laminar) case, while solid lines 
illustrate the turbulent case with $\delta_R=40\%$ and different $\ell$ 
(color code as in Fig.~2). In the left panel, solid lines refer to the 
thermal pressure, while dotted ones refer to the total pressure.} 
\end{figure*}

The direct and indirect effects of $\dot M$ dwindling combine into the 
accretion rate scaling $\dot M\propto v_1\,M/R\propto v_1^3/\Delta\phi$; 
whence the infall speed follows
\begin{equation}
v_1\propto \dot{M}^{1/3}\, (\Delta\phi)^{1/3}~.
\end{equation}
This is indeed reduced strongly during the late development of the outskirts 
when both $\dot M\propto d/\epsilon$ subsides and 
$\Delta\phi\propto\epsilon^{-2}$ lowers, so that the overall scaling 
$v_1\propto d^{1/3}/ \epsilon$ applies for $\epsilon>1$; specifically, as 
$\epsilon$ increases from $1$ to $1.5$ and then to $3$, the prevailing Mach 
numbers decline from $\mathcal{M}^2\approx 10$ to $6$ and then to $3$. 

Such values are consistent with the Mach number distributions at low 
$z$ sliced for flows of preheated gas into the cluster, as found by numerical 
simulations (e.g., \citealp{Ryu2003}; \citealp{Skillman2008}, see their 
Fig.~6 and Sect.~4; see also \citealp{Vazza2010}).

\subsection{Weakening shocks and entropy demise}

With $v_1$ lowering toward transonic values the shock strength will 
eventually \emph{weaken}. We recall \citep[see Appendix B in][]{Lapi2005} 
that the Rankine-Hugoniot conservation conditions for a standing shock give 
the postshock temperatures and densities in the form 
\begin{equation}
{T_2\over T_1} = {7\over 8} + {5\over 16}\, \mathcal{M}^2 - {3\over 16}\, 
{1\over \mathcal{M}^2}~~~~~,~~~~~{n_2\over n_1}={4\,\mathcal{M}^2\over 
3+\mathcal{M}^2}~.
\end{equation}
As the Mach number decreases from values $\mathcal{M}^2>3$ (defining strong 
shocks) to transonic $\mathcal{M}^2\approx 1$, the temperature ranges widely 
from $k_B T_2\simeq 3\,\mu m_p\,v_1^2/16$ to $T_2 \simeq  T_1$, while the 
density varies mildly from $n_2\simeq 4\,n_1$ to $n_2\simeq n_1$. 
Correspondingly, the shock-generated entropy $k_2\equiv k_BT_2/n_2^{2/3}$ 
drops from values $5\,\mathcal{M}^2\,k_1/4^{2/3}\,16\sim$ some $10^3$ keV 
cm$^2$ typical of strong shocks to intergalactic values $k_1=k_B 
T_1/n_1^{2/3}\lesssim 10^2$ keV cm$^2$.

By the same token, the entropy outer ramp will abate. Its slope at 
the boundary has been derived by \citet{Cavaliere2009} from the jumps at the 
shock and the adjoining HE recalled in Sect.~1, to read 
\begin{equation}
a_R = {\mathcal{A}\over 2}\, \left(5 - b_R\right)~;
\end{equation}
here $\mathcal{A} = 4\, (1+\epsilon/d)/[5+2\,(1+\epsilon/d)]\approx 1$ 
applies as long as $\epsilon\approx 1$ and $d\approx 2/3$ hold.

However, the main dependence of $a_R$ on $\epsilon$ and $d$ is encased into 
$b_R\equiv \mu m_p\, v_R^2/k_B T_R$ through the boundary temperature. We have 
just seen that $T_2\propto v_1^2\propto v_R^2\, \Delta\phi$ holds as long as 
strong shock conditions apply. If so, $b_R\propto 
1/\Delta\phi\propto\epsilon^2$ would rise fast as $\epsilon$ exceeds $1$; but 
$v_1^2\propto d^{2/3}/\epsilon^2$ is meanwhile reduced and the 
virial accretion shocks weakened, slowing down the net growth of 
$b_R$. The overall result is that $b_R$ grows from the standard value $2.7$ 
to about $5$ and then to about $8$ as $\epsilon$ increases from $1$ to $1.5$ 
and then to $3$. 

After Eq.~(5), increasing values of $b_R$ cause $a_R$ to decrease, 
and imply progressive saturation and even a decline of the entropy produced 
at the boundary. Specifically, as $\epsilon$ increases from $1$ to $1.5$ and 
then to $3$, the entropy slope $a_R$ goes from standard value $1.1$ to about 
$0$ and then down to negative values around $-2$.

To tackle the issue, recall that while the cluster outskirts develop to the 
currently observed radius $R$, the ICP is adiabatically compressed and 
settles into the DM potential well. Then the specific entropy $k(r)$ 
stratifies shell by shell, leading to a running slope $a(r)=a_R$ that retains 
the sequence of the values set at the time of deposition 
\citep[see][]{Tozzi2001, Lapi2005}. As a result, on moving out from the early 
cluster body to the outskirts currently building up by secular accretion, the 
whole slope $a(r)$ of the outer ramp decreases, and $k(r)\propto r^{a(r)}$ 
flattens or even bends over. 

To describe this behavior, in \citet{Lapi2010} we used an entropy slope 
$a(r)=a-a'\,(r-r_b)$ smoothly decreasing from the body value $a\approx 1.1$. 
We found a value $a\approx 0$ at $r\approx R/2$ and values $a\approx -2$ at 
$R$ (as illustrated in Fig.~1 of \citealp{Lapi2010}), consistently with the 
data by \citet{Bautz2009} and \citet{George2009}. 

\subsection{Onset of turbulence}

Here we show that such an entropy demise due to $\dot{M}$ dwindling will 
arise \emph{together} with the onset of turbulence triggered by shock 
weakening. In fact, the latter causes the postshock speeds 
\begin{equation}
{v_2\over v_1} = {n_1\over n_2}={1 \over 4}\, + {3\over 4}\, 
{1\over\mathcal{M}^2}~, 
\end{equation}
to grow from values around $1/4$ and approach $1$. So with $v_1\propto 
\epsilon^{-1}$ lowering sharply while $c_1$ varies as $(1+z)$ or less, the  
kinetic energy ratio is \emph{enhanced} relative to the strong shock value 
$6.3\%$, see Sect.~2.1. For example, as $\epsilon$ grows from $1$ to $1.5$ 
and then to $3-5$, the ratio $v_2^2/v_1^2$ increases from $10\%$ to $14\%$ 
and then to $25-39\%$. Note that the ratio $v_2^2/c_2^2$ of the residual bulk 
to the sound's speed past the shock also increases for decreasing 
$\mathcal{M}^2$; in particular, it goes from $25\%$ to $50\%$ as 
$\mathcal{M}^2$ ranges from $1$ to $3-5$. On average, for a CC cluster we 
find the condition $\mathcal{M}^2 \propto 
d^{1/3}/\langle\epsilon(z)\rangle\,(1+z)\lesssim 3$ for shock weakening to be 
met at redshifts $z\lesssim 0.3$, on using for the average $\langle\epsilon 
\rangle$ the values given by \citet{Lapi2009} in their Fig.~6. In a nutshell, 
combining Eqs.~(4) and (6) yields an \emph{inverse} relation of 
$v_2^2/v_1^2$ with $k_2/k_1$. 

These postshock flows provide bounds to the energy level of \emph{subsonic} 
turbulent motions that is driven by smooth accretion in relaxed clusters; 
minor, intermittent and localized contributions may be added by the 
complementary clumpy component of the accretion, recently re-calibrated to 
less than $30\%$ in the outskirts of relaxed halos 
\citep[see][]{Fakhouri2010, Genel2010, Wang2010}. Our bounds actually 
constitute fair estimates for the amplitudes of outer turbulent energy at 
$r\sim R$, as shown by similar values obtained both observationally 
\citep[see][]{Mahdavi2008, Zhang2010} and in many numerical simulations from 
\citet{Evrard1990} to \citet{Lau2009}. 

\section{Modeling the turbulent support}

Turbulent motions start at the virial radius $R$ with comparable coherence 
lengths $L\sim R/2$, set in relaxed CC clusters by the pressure scale height 
or by shock segmentation (see \citealp{Iapichino2008}, \citealp{Ryu2008}, 
\citealp{Pfrommer2010}, \citealp{Vazza2010}). Then they fragment downstream 
into a dispersive cascade over the `inertial range', to sizes $\ell$ where 
dissipation begins after the classic picture by \citet{Kolmogorov1941}, 
\citet{Obukhov1941} and \citet{Monin1965}. In the ICP context the dissipation 
scale (equivalently to the classic Reynolds' scaling) writes $\ell\sim 
(c_2/\tilde{v})^{3/4}\, \lambda_{\rm pp}\, (L/\lambda_{\rm pp})^{1/4}$ in 
terms of the ion collisional mean free path $\lambda_{\rm pp}$ and of the 
ratio $\tilde{v}/c_2$ of the turbulent rms speed to the sound's, see 
\citet{Inogamov2003}. For subsonic turbulence with $\tilde{v}/c_2\lesssim 
1/3$ (see direct observational bounds by \citealp{Schucker2004}, 
\citealp{Sanders2010}, and references therein) the relevant scale $\ell$ 
exceeds somewhat $\lambda_{\rm pp}\sim 10^2$ kpc. 
 
In the presence of outer magnetic fields $B\lesssim 10^{-1}\,\mu$G 
(see \citealp{Bonafede2010}, \citealp{Pfrommer2010}; also \citealp{Ryu2008}) 
the key if quantitatively debated feature is their degree of tangling; then 
then the effective mean free path is provided by the coherence scale of the 
field, somewhat larger than $\lambda_{\rm pp}$ (see discussions by 
\citealp{Narayan2001}, \citealp{Inogamov2003}, \citealp{Govoni2006}, 
\citealp{Brunetti2007}).

Clearly the above estimates provide useful guidelines, but need theoretical 
modeling and observational probing. Both issues are addressed next. 

Since turbulent motions contribute to the pressure \citep[see][]{Landau1959} 
to sustains HE, our modeling is focused on the ratio $\delta(r)\equiv p_{\rm 
nth}/p_{\rm th}$ of turbulent to thermal pressure (or, equivalently, on the 
ratio $\delta/(1+\delta)$ of turbulent to total pressure), with radial shape 
decaying on the \emph{scale} $\ell$ from the \emph{boundary} value 
$\delta_R$. 

The total pressure is conveniently and generally written as $p_{\rm th}(r)\, 
[1+\delta(r)]$, while the thermal component is still expressed as $p_{\rm 
th}(r)\propto k(r)\, n^{5/3}(r)$. With this addition, we proceed to solve the 
HE equation just along the steps leading to Eq.~(1), and now find the 
temperature profile in the form 
\begin{eqnarray}
\nonumber{T(r)\over T_R}&=&\left[{k(r)\over k_R}\right]^{3/5}\,
\left[{1+\delta_R\over 1+\delta(r)}\right]^{2/5}\,\left\{1+{2\over 5}\, 
{b_R\over 1+\delta_R}\, \times \right.\\
&\times& \left.\int^R_r{\mathrm{d} x\over x}\,{v_c^2(x)\over v_R^2}\,
\left[{k_R\over k(x)}\right]^{3/5}\, \left[{1+\delta_R\over
1+\delta(x)}\right]^{3/5}\right\}~,
\end{eqnarray}
which extends Eq.~(1) for $\delta>0$. Again, $n(r)$ is linked to $T(r)$ by 
$n=[k_B T/k]^{3/2}$. 

The actual temperature at the boundary is now lowered to 
$T_R=T_2/(1+\delta_R)$. This is seen from Rankine-Hugoniot conditions in the 
presence of turbulent pressure; the latter obviously implies the term 
$p_2\,(1+\delta_R)$ to be added on the right hand side of the stress balance, 
and the corresponding one $5\,p_2 \, (1+\delta_R)\,v_2/2$ in the energy flow 
\citep[see Eqs.~B1 in][]{Lapi2005}.

In our numerical computations that follow we will adopt for fully 
developed turbulence the simple functional shape (rather than a constant 
$\delta$ as used by \citealp{Bode2009}) 
\begin{equation}
\delta(r)=\delta_R\, e^{-(R-r)^2/\ell^2}~,
\end{equation}
which decays on the \emph{scale} $\ell$ inward of a round maximum, a smoothed 
out representation of the inertial range. This is illustrated in Fig.~2 for 
three cases with $\ell=0.9\, R$ (blue), $\ell=0.5\, R$ (red) and $\ell=0.2\, 
R$ (green). Note that the runs $\delta(r)$ we adopt are consistent with those 
recently indicated by numerical simulations (e.g., \citealp{Lau2009}).

\section{Results}

We illustrate in Fig.~3 the resulting temperature and pressure profiles. 
Relative to pure thermal HE with $\delta_R = 0$, in the outskirts the thermal 
pressure lowers since it is helped by turbulent motions in sustaining the 
equilibrium, while the pressure run is moderately steeper; at the center the 
pressure is mainly contributed by the thermal component. 

\begin{figure}
\begin{minipage}{\linewidth}
\centering
\includegraphics[width=8cm]{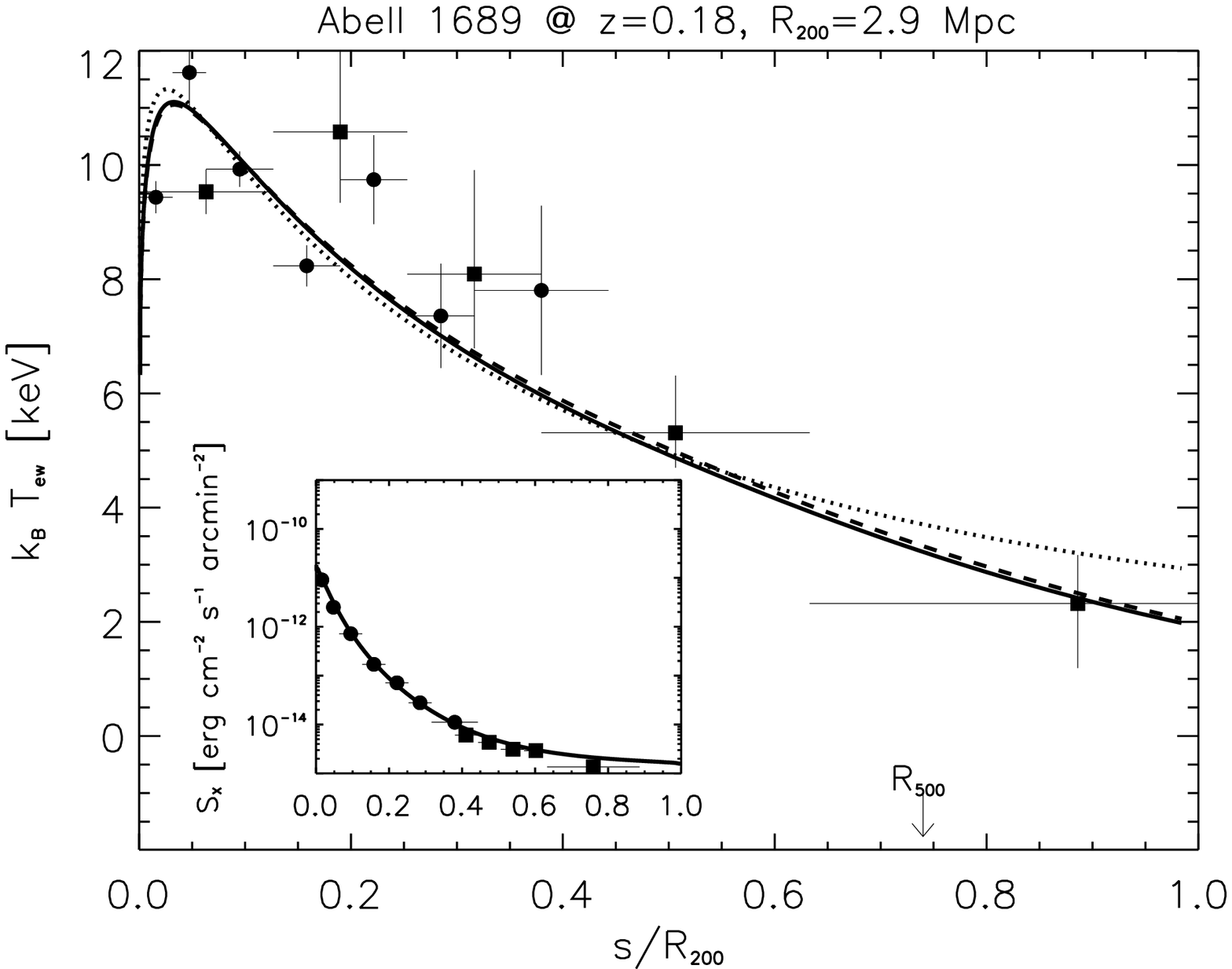}
\vspace{0.5cm}
\end{minipage}
\begin{minipage}{\linewidth}
\centering
\includegraphics[width=8cm]{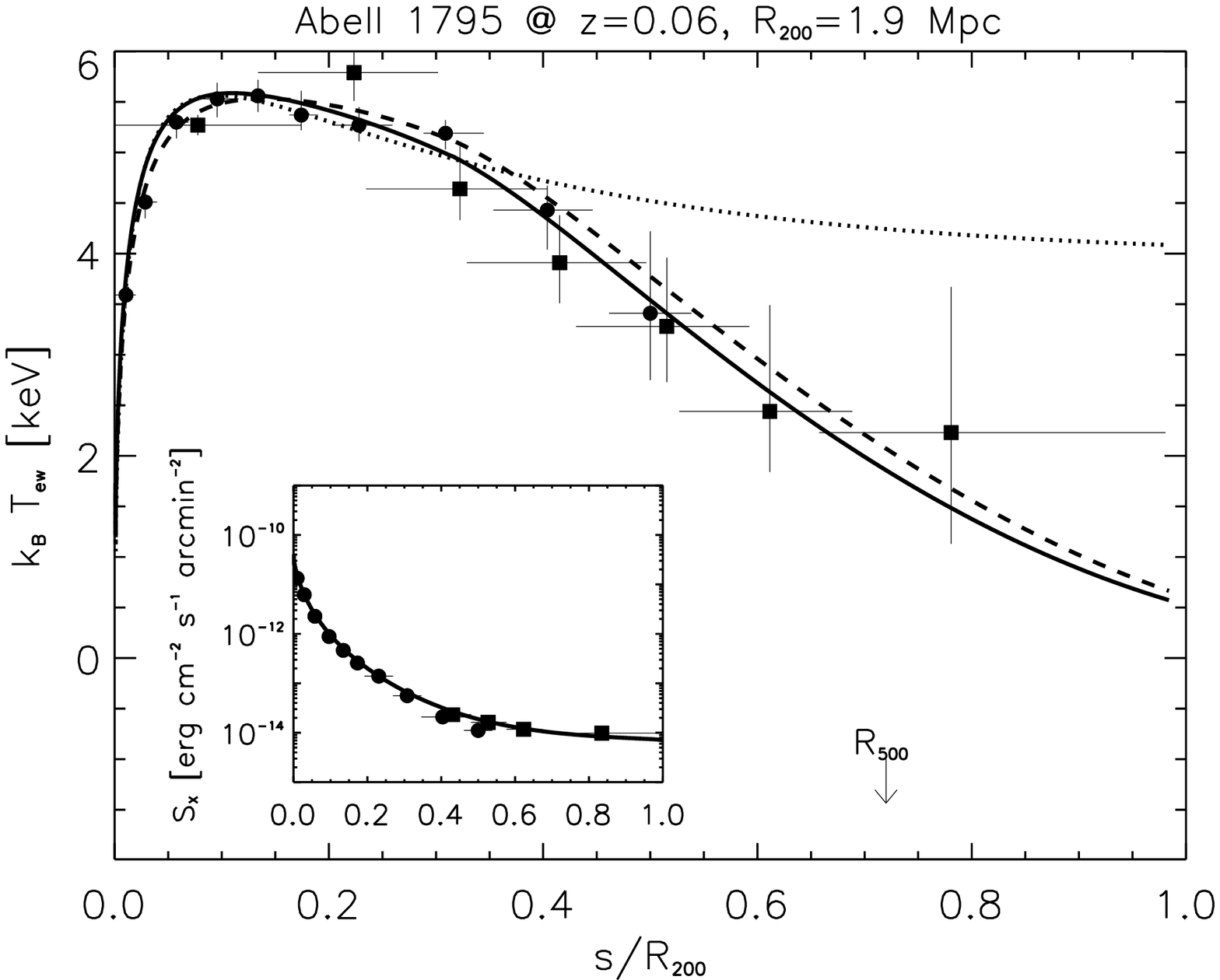}
\vspace{0.5cm}
\end{minipage}
\caption{Profiles of projected X-ray temperature (brightness in the 
insets) for the CC clusters A1689 (top) and A1795 (bottom). Data are from 
\citet{Snowden2008} with \textsl{XMM-Newton} (circles), and from 
\citet{Bautz2009} and \citet{Kawaharada2010} with \textsl{Suzaku} (squares). 
The solid lines represent our bestfits with the SM extended to include 
turbulence after Eqs.~(7) and (8) with $\delta_R=40\%$ and $\ell=0.5\, R$.
The dashed ones illustrate the outcomes in the absence of turbulence, 
but still with entropy decreasing outwards as in \citet{Lapi2010}. For 
comparison, the dotted lines illustrate the case with uniform entropy 
slope.} 
\end{figure}

As to the temperature (with the classic CC shape), it is seen that the 
variations are mild, and primarily stem from the reduction of $T_R/T_2$ by 
the factor $1+\delta_R=1.4$ discussed in Sect.~3. Thus we recognize the 
saturation or bending over of entropy on scales $r\lesssim R/2$ (see Sect.~2) 
to constitute the primary cause for the steep temperature profiles observed 
by \textsl{Suzaku}; these drop by factors around $10$ from the peak to the 
outer boundary in the low-$z$ clusters like A1795 and PKS0745-191. 

This view is confirmed by Fig.~4, where we illustrate our bestfits 
with the SM to the temperature and surface brightness profiles for the two 
clusters A1795 (north sector) and A1689 (azimutally averaged), one at a low 
and the other at a relatively high $z$. The figure shows that the SM with an 
entropy slope decreasing through the outskirts from standard values $a\approx 
1.1$ at $r<R/2$ to $a\approx -1.8$ at $R$ (just as expected in Sect.~2.2) 
fits the temperature profile of the low-$z$ cluster A1795 much better than 
the case with uniform slope $a\approx 1.1$. As expected, the difference is 
barely discernible for the relatively high-$z$ cluster A1689. Both fits are 
only mildly affected on including in the SM the turbulent support. Note that 
the linked $n(r)$ profiles flatten out to enhanced brightness landings (cf. 
insets in Fig.~4), a simple warning of interesting temperature and turbulence 
distributions.

A far reaching \emph{consequence} of turbulent support shows up (see Fig.~5, 
left) in the reconstructions of DM masses from X-ray observables based on 
reversing the thermal HE equation \citep[cf.][pag. 92]{Sarazin1988}. 
It is seen that the mass reconstructed on ignoring the 
non-thermalized component deviates from the true mass by different amounts, 
and for small values of the dissipation scale $\ell$ may even have a 
non-monotonic behavior, not unlike the results by \citet{Kawaharada2010} in 
three sectors of A1689 (cf. their Fig.~8). In all cases the reconstructed 
mass within $R$ is negatively \emph{biased} by $20-30\%$ relative to the true 
one. Accounting for such a bias constitutes a key point to resolve the 
tension between weak lensing and X-ray masses \citep[e.g.,][]{Nagai2007, 
Lau2009, Meneghetti2010}, and in deriving precise cosmological parameters 
from statistics of cluster masses via the fast X-ray observations 
\citep[see][]{Vikhlinin2009}. 

The above picture may be double-checked in individual clusters on directly 
gauging with the SZ scattering how the electron pressures are lowered in the 
presence of outer turbulence. In Fig.~5 (right) it is seen that along l.o.s. 
running mainly into the outskirts we expect the latter to be lower by 
$20-30\%$ relative to the pure thermal case. This is stronger by a factor 
about $5$ than the effects from delayed equipartition between electron and 
ion temperatures downstream the shock over the relevant mean free path. A 
corresponding reduction is implied in the requirements for sensitivity and/or 
observation times with upcoming sophisticated instruments like \textit{ALMA} 
\citep[see][]{Wong2009}. Meanwhile, statistical evidence of SZ reductions has 
been extracted from stacked data by \citet{Komatsu2010}. 

\begin{figure*}
\centering
\includegraphics[width=15cm]{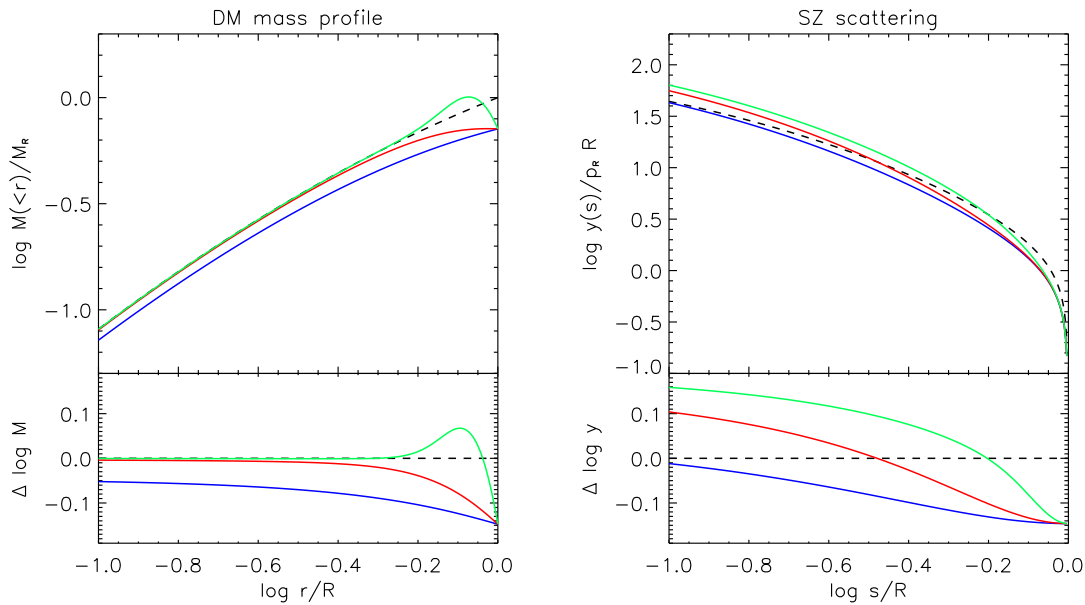}
\caption{Left: profile of DM mass; the dashed line illustrates the true mass, 
while the solid lines illustrate that reconstructed from X-ray observables on 
assuming pure thermal HE. Right: Projected profile of SZ scattering; the 
dashed line refer to the pure thermal case, while solid lines refer to the 
turbulent case. In both panels $\delta_R=40\%$ and different values of $\ell$ 
(color code as in Fig.~2) are adopted.} 
\end{figure*}

\section{Discussion and conclusions}

To answer the issue raised at the end of Sect.~1, we have investigated the 
connection of ICP turbulence in cluster outskirts with virial accretion 
shocks; among the variety of shocks found in numerical simulations, these are 
best amenable to a simple treatment, that may also shed light on more complex 
conditions. 

A result to stress is the \emph{inverse} nature of the connection that 
follows from combining the key Eqs.~(4) and (6), namely, as the Mach number 
$\mathcal{M}$ of the shock decreases the postshock entropy $k_2\propto 
T_2/n_2^{2/3}$ lowers with $\mathcal{M}^2$ while the residual bulk energy 
$v_2^2/v_1^2$ rises with $\mathcal{M}^{-2}$. In other words, 
\emph{saturation} of entropy production and \emph{increasing} bulk flows to 
drive turbulence occur \emph{together}. 

To pinpoint when this is bound to occur, we hinge upon the quantity $\dot M\, 
v_1^2\, (v_2/v_1)^2\propto d^{5/3}\,\epsilon^{-3}\, (v_2/v_1)^2$ including 
three factors: the infall speed $v_1^2\propto d^{2/3}/\epsilon^{2}$, see 
Eq.~(3); the accretion rate $\dot M\propto d/\epsilon$; and the residual 
kinetic energy ratio $v_2^2/v_1^2$, see Eq.~(6) and discussion thereafter. 
Such a quantity depends \emph{strongly} on $d$ and even more on $\epsilon$, 
which render the effects on the outskirts growth of the cosmological 
expansion accelerated by the Dark Energy, and of the declining shape of the 
initial DM perturbation wings, respectively.

This leads us to predict that the turbulent support starts to increase on 
average for $z<1/2$ during the late development of the cluster outskirts when 
$d\approx 1/2$ and $\epsilon>1$ apply. Then turbulence briskly \emph{rises} 
at $z\lesssim 0.3$ when the shocks become transonic. 

On the other hand, environmental (even anisotropic) variance is introduced 
when the residual energy flow $\dot M\,v_2^2$, strongly dependent on 
$\epsilon$, is modulated by the adjacent filament/void structure. At very low 
$z$ a high level of thermal support will persist in sectors adjacent to 
filaments, as is apparently the case with A1795 and PKS0745-191 
\citep[see][]{Bautz2009, George2009}. By the same token, values $\epsilon>1$ 
will prevail in cluster sectors facing a void, causing there an early onset 
of turbulence given sufficient $\dot{M}$. Meanwhile, in neighboring sectors 
facing a filament values $\epsilon\approx 1$ still hold and the thermal 
support may still prevail, a condition that apparently applies to A1689 
\citep[see][]{Kawaharada2010, Molnar2010}. 

With the closer focus provided by the Supermodel we find that as the 
shocks weaken not only the boundary entropy production saturates, but also 
the whole outer entropy distribution $k(r)$ is to flatten and the temperature 
profiles $T(r)$ to become \emph{steeper}, consistently with the observations. 
Eventually, as the inflow approaches the transonic regime, $k(r)$ tends to 
bend over and $T(r)$ to steepen somewhat further.

But then subsonic turbulence arises at the boundary; its inner 
distribution follows the classic picture of the turbulent motion 
fragmentation, with a debated scale for its final dissipation. This demands 
closer probing, that we have tackled with the fast, analytic tool provided by 
the Supermodel. The present data from joint X-ray and weak lensing 
observations \citep[see][]{Kawaharada2010, Molnar2010, Zhang2010} concur with 
simulations \citep[see][]{Lau2009} to show no evidence for a dissipation 
scale much shorter than $\sim 10^2$ kpc, conceivably set by tangling of the 
magnetic field \citep[see discussions by][]{Narayan2001, Brunetti2007}.

To summarize, in the outskirts of relaxed clusters we find:

\noindent $\bullet$ Turbulence is related to weakening shocks at late cosmic 
epochs $z<0.3$, when saturation of entropy production causes steep 
temperature profiles. 

\noindent $\bullet$ Turbulent excess pressures $\delta_R=p_{\rm 
nth}/p_{\rm th}$ up to $40\%$ arise at the boundary, declining inwards on 
scales $\ell\sim 100$ kpc.

\noindent $\bullet$ The overall masses derived from X rays are 
necessarily biased low down to $20-30\%$ when such a turbulent support is 
ignored. 

These findings are consistent with the current observational and 
numerical data. Moreover, we predict:

\noindent $\bullet$ Variance concerning steep $T(r)$ and turbulent support in 
cluster sectors will be correlated with the filament modulation of the 
adjacent environment. 

\noindent $\bullet$ The SZ scattering will be considerably lowered relative 
to the pure thermal case, along l.o.s. running mainly into the outskirts or 
their sectors where X rays concur with lensing data in signaling turbulent 
support. 

We stress that the SM provides a fast tool to represent and probe conditions 
of smooth inflows that prevail in the outskirts of relaxed CC clusters, away 
from mergers that scar the NCCs and constitute the realm of detailed but 
time-consuming numerical simulations. 

A final comment concerns the giant radiohalos observed at the centers 
of several clusters to emit synchrotron radiation; these suggest that 
non-thermal support may be contributed by a mixture of magnetic field and 
relativistic particles accelerated by shocks and turbulence due to mergers 
\citep[see][]{Brunetti2007b, Biermann2009, Brunetti2009}. These processes, 
with their limited acceleration efficiency and short persistence, are 
apparently widespread in NCC clusters, mainly at $z>0.2$; so they have minimal 
superposition or interference with the substantial, low-$z$, long-lived, 
outer turbulent component concerning mainly CCs, that we have addressed here. 

The picture we pursue envisages the infall kinetic energy to thermalize along 
two channels. First, supersonic inflows achieve thermalization at the shock 
transition via a sharp jump (or a few jumps within a layer of limited 
thickness, see \citealt{Lapi2005}). Second, the subsonic turbulent motions 
left over downstream the shocks, particularly downstream weak shocks, are 
dispersed into a cascade of many effective degrees of freedom 
\citep[see][]{Landau1959}, down to scales where dissipation becomes 
effective. The channels' branching ratio shifts toward the latter when the 
shock weaken, for $z\lesssim 0.3$; meanwhile, turbulence concurs with thermal 
pressure to support the ICP equilibrium in the outskirts. The picture 
substantiates the following formal remark: Eq.~(7) corresponds to Eq.~(1) for 
the variable $T\,(1+ \delta)$ in terms of the extended entropy 
$k\,(1+\delta)$. 

This picture of onset and development of outer ICP turbulence may be fruitful 
in other contexts, in particular for shocks and turbulence in wakes 
around mergers. So it warrants close \emph{modeling} with our fast yet 
precise analytic SM, to \emph{probe} over wide cluster samples in a range of 
$z$ the key turbulence features: amplitude and scale. 

\begin{acknowledgements}
Work supported by ASI and INAF. We thank our referee for helpful comments and 
suggestions. AL thanks SISSA and INAF-OATS for warm hospitality. 
\end{acknowledgements}

\end{document}